%% file: rem1d.tex
\documentclass[aps,pre,preprint,showpacs,superscriptaddress]{revtex4}

\usepackage{amsfonts,amsmath,amssymb}
\usepackage{graphicx,paralist}

\newcommand{\s}{\sigma}
\renewcommand{\P}{\mathbb{P}}

\renewcommand{\O}{\mathcal{O}}
\newcommand{\eqd}{\overset{\rm d}{=}}

\begin{document}
\date{\today}

\title{Mosaic length and finite interaction-range effects in a one
  dimensional random energy model}

\author{S.~Franz}
\affiliation{Universit\'e Paris-Sud, LPTMS, UMR8626, B\^at.\ 100,
  91405 Orsay cedex, France}

\author{G.~Parisi} 
\affiliation{Dipartimento di Fisica, INFM-CNR SMC and INFN sez.\ di
  Roma1, Universit\`a di Roma ``La Sapienza'', P.le A. Moro 2, Roma
  00185, Italy}

\author{F.~Ricci-Tersenghi} 
\affiliation{Dipartimento di Fisica, INFM-CNR SMC and INFN sez.\ di
  Roma1, Universit\`a di Roma ``La Sapienza'', P.le A. Moro 2, Roma
  00185, Italy}

\pacs{05.20.-y (Classical statistical mechanics),
64.70.Pf (Glass transitions),
75.10Nr (Spin-glass and other random models)}

\begin{abstract}
  In this paper we study finite interaction range corrections to the
  mosaic picture of the glass transition as emerges from the study of
  the Kac limit of large interaction range for disordered models.  To
  this aim we consider point to set correlation functions, or
  overlaps, in a one dimensional random energy model as a function of
  the range of interaction. In the Kac limit, the mosaic length
  defines a sharp first order transition separating a high overlap
  phase from a low overlap one.  Correspondingly we find that overlap
  curves as a function of the window size and different finite
  interaction ranges cross roughly at the mosaic lenght. Nonetheless
  we find very slow convergence to the Kac limit and we discuss why
  this could be a problem for measuring the mosaic lenght in realistic
  models.
\end{abstract}

\maketitle

\section{Introduction}

The paradigm of `random first order transition' or one-step replica
symmetry breaking (1RSB) theory, provides an elegant framework to
conceptualize the phenomenology of liquids approaching the glass
transition \cite{1rsb}. Unfortunatly, this scenario is strongly based
on mean-field models \cite{ktw} and mean-field-like approximations to
liquid theories \cite{mp} and cannot be taken literally in the
application to real system. The main node that has to be untied to
establish the 1RSB scenario as a convincing theory for real materials,
is how mean-field theory should be adapted and modified to take into
account the finite range of interactions. Though a fundamental theory
of glassy systems in finite dimension is presently lacking, proposals
have been made that modify minimally mean field scenario to take into
account the finite interaction range. In ref.~\cite{ktw} Kirkpatrick,
Thiurmialai and Wolynes developed a phenomenological theory, known as
`mosaic picture', where it is postulated the existence of a coherence
length, that grows on lowering the temperature. Below that length the
system behaves essentially as a mean-field glass, while it would
cross-over to liquid behavior at larger scales. It results a theory
where relaxation is dominated by activated processes stemming from the
competition between interface tension and a bulk configurational
entropy. The mosaic picture has been recently revived and deeply
clarified by Biroli and Bouchaud~\cite{BB}, who showed that while
usual (point-to-point) correlation functions are insensitive to the
possible growth of the coherence mosaic length, it is possible to
define different ``point-to-set'' correlation functions, able to
reveal the growth of the mosaic length. In turn, the mosaic length has
been related to the relaxation time of ordinary, time dependent
correlation functions \cite{MonSem}.  These papers prompted on one
side numerical simulations on kinetically constrained glasses
\cite{garrahan} and on realistic glassy models \cite{Cavagna}, on the
other to theoretical calculations for models on trees and under the
Kac limit \cite{FM}. These last models are the natural starting point
for understanding the mosaic picture, since their local properties are
well described by mean-field theory \cite{kac}.  In \cite{FM} the
study of point to set correlation function has allowed to derive a
detailed picture relating the relaxation in the Mode Coupling regime
for $T>T_d$ to the one in the mosaic regime for $T<T_d$. The
calculation, supposedly exact, concern the behavior of disorderd
glasses in the Kac limit. In order to understand its relevance for
short range systems, it is necessary to study the properties of
convergence to the Kac limit for finite interaction range. It has been
found in ref.~\cite{Cavagna} that in standard Lennard-Jones
supercooled liquids, the transition from high to small overlap as a
function of the box size is much smoother than one would expect from
the mosaic picture. This poses the question of what behavior one
should expect when the range of interaction is not large.

In this paper, we address this question in a minimalistic finite
dimensional model displaying 1RSB behavior in the Kac limit. The model
is a one dimensional version of the Random Energy model \cite{REM1}
extensively studied in the context of stochastic models for reaction
diffusion equations and evolving populations \cite{evol}. This has two
main advantages: on one hand the Kac limit can be studied directly by
probabilistic arguments, without having to resort to replicas or
cavity techniques, on the other the model for finite interaction range
can be studied exactly by transfer matrices.

A recent paper addresses the problem of finite range corrections to
the mosaic picture in a related one dimensional XORSAT model
\cite{Montanari}. That paper concerns the zero temperature limit,
while we concentrate on finite temperature properties.

The organization of the paper is the following: in section~\ref{sec1}
we define the model. Section~\ref{sec2} is devoted to the definition
of the point to set correlation we study. In section~\ref{sec3} we
discuss theoretical approaches to the computation of this quantity. In
section~\ref{sec4} we discuss the results of exact computations with
transfer matrices. Finally we draw our conclusions.

\section{The model}\label{sec1}

In order to compare the behavior of finite range interaction systems
with mean-field theories we need a model with variable interaction
range which is well suited for numerical analysis.  We decided to
consider a 1D version of the Random Energy Model (REM) \cite{REM1}
introduced in the first of ref. \cite{evol}. This consists in a line
of $m\,L$ Ising spins, divided in $L$ groups of $m$ spins such that
only neighbouring groups of spins interact (thus leading to an
interaction range of $2m$).

For each group $i=1,...,L$ we define a state variable $\s_i$ taking
values $1,...,2^m$.  In the variables $\{\s_i\}$ the interactions are
restricted to nearest neighbors. The Hamiltonian of the system is
\begin{equation}
H(\vec\sigma) = \sum_{i=0}^{L-1} E_i(\sigma_i,\sigma_{i+1})\;.
\end{equation}
For each link the $2^{2m}$ interaction energies $E_i(\sigma,\tau)$ are
quenched random variables extracted from a Gaussian distribution of
zero mean and variance
\begin{equation}
\overline{E^2} = m/2\;.
\end{equation}

We have considered fixed boundary conditions on the left side (in
$i=0$), defining $\s_0=1$, and open boundary conditions on the right
side (in $i=L$). In this way we minimize the computational effort
needed to compute the free-energy $Z_L$, which is expressed as
$Z_L=\sum_\s Z_L(\s)$, where $Z_L(\s)$ is given by the recursion
relation
\begin{equation}
Z_{\ell+1}(\s)=\sum_{\tau=1}^{2^m} Z_\ell(\tau) e^{-\beta
E_\ell(\tau,\s)}\;,
\end{equation}
with $Z_0=1$ and $\beta=1/T$.  Computing $Z_L$ thus requires
$\O(L\,2^{2m})$ operations.

In the $m\to\infty$ limit, the thermodynamics is simple: the
correlations between the energy level implied by the one dimensional
structure are negligible and, independently of $L$, the free-energy
coincides with the one of a REM with $2^{mL}$ states and energies
distributed according to $P(E)\propto \exp(-E^2/mL)$:
\begin{equation}
F=\lim_{m\to\infty}-\frac{T}{mL}\log Z_L=\left\{
\begin{array}{cl} 
-\frac{\beta}{4}-T\log(2) & \quad T>T_c\;,\\
-\sqrt{\log(2)} & \quad T\le T_c\;,
\end{array}
\right.
\end{equation}
with $T_c=\left(2\sqrt{\log(2)}\right)^{-1}$.

\section{The observables}\label{sec2}

Here we define the correlation functions of interest, allowing us to
detect a growing static length. These are built with the aid of a
suitably chosen reference configuration $\{\s_i^*\}_{i=1,...,L}$, to
which one fixes the system outside a window with $\ell$ sites located
around the center of the system.  For convenience we renumber
$1,...,\ell$ the sites in the central window.  Inside the window the
system is at thermal equilibrium. We investigate the correlation among
typical in-window configurations $\vec\s$ with $\vec\s^*$ to see
whether a characteristic length exist $\ell_c$ such that for window
sizes $\ell<\ell_c$, $\vec\s \simeq \vec\s^*$ inside the window,
while, for $\ell > \ell_c$, $\vec\s$ and $\vec\s^*$ are uncorrelated.

As detailed in the following, in order to sharpen the transition from
correlated to uncorrelated behavior we decided to fix the reference
configuration $\vec\s^*$ always to the ground state.  We then study
the thermodynamics of a system which is fixed to the reference
configuration outside a window of size $\ell$:
\begin{equation}
\s_i = \s_i^* \qquad \forall i < 1 \text{ and } \forall i > \ell\;.
\end{equation}
The system has then fixed boundaries and $\ell$ free variables,
$\{\s_i\}_{i=1,...,\ell}$.  Within the window, we can define its
overlap with respect to the reference configuration as
\begin{equation}
q(\vec\s, \vec\s^*) \equiv \frac{1}{\ell} \sum_{i=1}^\ell
\delta(\s_i,\s_i^*)\;.
\end{equation}
Notice that our point-to-set correlation function differs from the one
defined in \cite{BB} and used subsequently which consists in choosing
$\sigma^*$ as a configuration thermalized at temperature $T$.

We need some observable estimating the similarity of the typical
configuration with respect to the reference one, and to this end we
introduce the following two quantities:
\begin{eqnarray}
p_0(\ell,\beta) &\equiv&
\frac{e^{-\beta\sum_{i=0}^\ell E_i(\s_i^*,\s_{i+1}^*)}}
{\sum_{\{\s_i\}_{i=1}^\ell}
e^{-\beta\sum_{i=0}^\ell E_i(\s_i,\s_{i+1})}}\;,\\
q_0(\ell,\beta) &\equiv&
\frac{\sum_{\{\s_i\}_{i=1}^\ell} q(\vec\s, \vec\s^*)
e^{-\beta\sum_{i=0}^\ell E_i(\s_i,\s_{i+1})}}
{\sum_{\{\s_i\}_{i=1}^\ell}
e^{-\beta\sum_{i=0}^\ell E_i(\s_i,\s_{i+1})}}\;,
\end{eqnarray}
where the denominator is the ``window partition function''.  The first
quantity, $p_0$, is the relative weight of the reference configuration
in the window partition function computed at inverse temperature
$\beta$, while the second quantity, $q_0$, is the mean overlap with
the reference configuration.  Both quantities still depend on the
quenched disorder and we compute their typical values by
$\log(p_{typ}) \equiv \overline{\log(p_0)}$ and $\log(q_{typ}) \equiv
\overline{\log(q_0)}$, where the overline stands for the average over
the quenched disorder.  We expect $\log(p_0)$ and $\log(q_0)$ to be
self-averaging, since their are related to free-energy differences.

\section{Theoretical analysis}\label{sec3}

In this section we address the problem of an analytic computation of
the correlation functions. We will first study exactly the asymptotic
long range limit $m\to\infty$. After that we will address the problem
of finite $m$ effects, that our numerical analysis below reveals to be
very large.

\subsection{The correlation functions for $m\to\infty$}

\begin{figure}
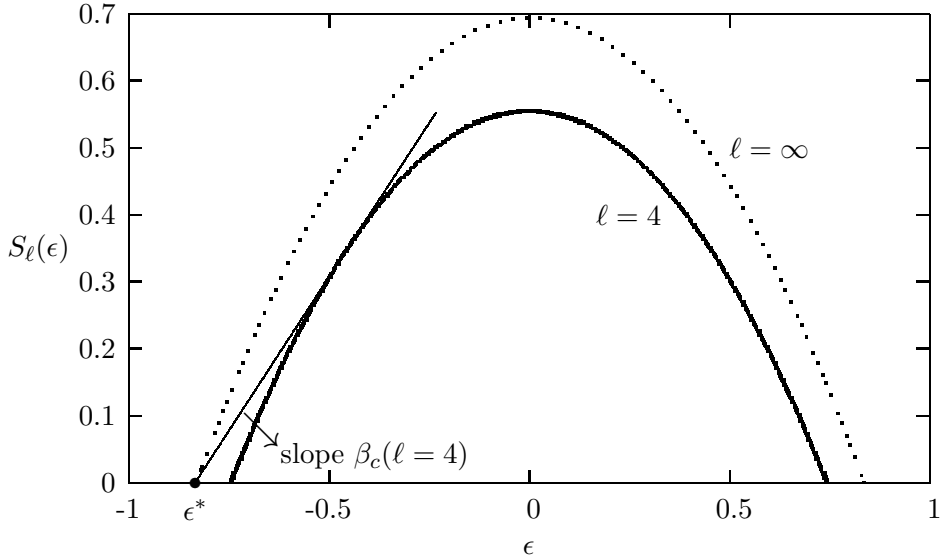

\include{parabole}
\caption{Microcanonical entropy for the full system ($\ell = \infty$,
  dotted line) and for a window of size $\ell=4$ (thick line).  A
  system prepared in the ground state of the full system (big dot) and
  constrained in a window of size $\ell = 4$ makes a first order
  transition at inverse temperature $\beta_c(\ell = 4)$ (the slope of
  the thin line).}
\label{fig:parabole}
\end{figure}

The infinite $m$ limit can be understood since in this limit the
correlations between the energy level due to the one-dimensional
structure of the model become negligible.  In this case, using this
independence approximation, we see that, besides the state $\vec\s^*$
of energy $E^*$, the window has $2^{m\ell}-1$ states with energies
distributed according to $P_\ell(E) \propto e^{-E^2/m(\ell+1)}$ (there
are $\ell$ sites and $\ell+1$ links!).  So that the average density of
states is
\begin{equation}
{\cal N}(E) \sim 2^{m \ell}e^{-E^2/m(\ell+1)}+\delta_{E,E^*}
\end{equation}
and the microcanonical entropy per link (divided by $m$) as a function
of the link energy $\epsilon=E/m(\ell+1)$ is
\begin{equation}
S_\ell(\epsilon)=\left\{
\begin{array}{cl}
-\epsilon^2+\frac{1}{1+1/\ell}\log(2) &
\quad |\epsilon|<\sqrt{\frac{1}{1+1/\ell}\log(2)}\\
0 & \quad \text{otherwise}
\end{array}
\right. 
\end{equation}
From this function the canonical thermodynamics can be derived.
Before doing that, few comments are in order: (a) The constrained
entropy is reduced by a constant term with respect to the
unconstrained case, given by the above formula with $\ell=\infty$ (see
Fig.~\ref{fig:parabole}). (b) The choice of the reference
configuration  $\vec\s^*$ as the ground
state has no effect on the other states: the same entropy would be
obtained for different choices of $\sigma^*$. Of course the window
thermodynamics and correlations would depend on the energy of
$\vec\s^*$.

If the state $\vec\s^*$ was absent, the free-energy per link would
read
\begin{equation}
{\tilde f}(\beta,\ell)=\left\{
\begin{array}{cl}
-\frac{\beta}{4}-\frac{T}{1+1/\ell}\log(2) &
\quad T > T_c\sqrt{1+1/\ell}\\
-\sqrt{\frac{1}{1+1/\ell}\log(2)} &
\quad T \le T_c\sqrt{1+1/\ell}.
\end{array}
\right.
\end{equation}
Including the state $\vec\s^*$ one therefore has
\begin{equation}
f(\beta,\ell)=\min\{\epsilon^*,{\tilde f}(\beta,\ell)\}\;,
\label{eq:free}
\end{equation}
with $\epsilon^*=-\sqrt{\log(2)}$, that is the ground state energy of
the $\ell=\infty$ system.  When the two terms in Eq.(\ref{eq:free})
are equal, a first order transition takes place (see
Fig.~\ref{fig:parabole}) at inverse temperature
\begin{equation}
\beta_c(\ell)=2 \sqrt{\log(2)} \left(1-\frac{1}{\sqrt{\ell+1}}\right)
= \beta_c \left(1-\frac{1}{\sqrt{\ell+1}}\right)\;,
\end{equation}
which in turns defines a temperature dependent critical length
\begin{equation}
\ell_c(\beta) = \frac{\beta(2\beta_c-\beta)}{(\beta_c-\beta)^2}\;,
\label{eq:ell_c}
\end{equation}
separating the confined regime $\ell<\ell_c$ where $p_{typ}=q_{typ}=1$
from the deconfined regime $\ell>\ell_c$ where $p_{typ}=q_{typ}=0$.

The size of the critical window diverges as expected at the critical
temperature, where the configurationl entropy vanishes.  We find that
this critical length is quardatic in the inverse of $T-T_c$; had we
chosen the reference state $\vec\s^*$ with a different rule, the
result would have been different.  For example a direct calculation
shows that choosing $\vec\s^*$
with Boltzmann probability at temperature $T$ implies a linear
critical length in $1/(T-T_c)$.

\begin{figure}
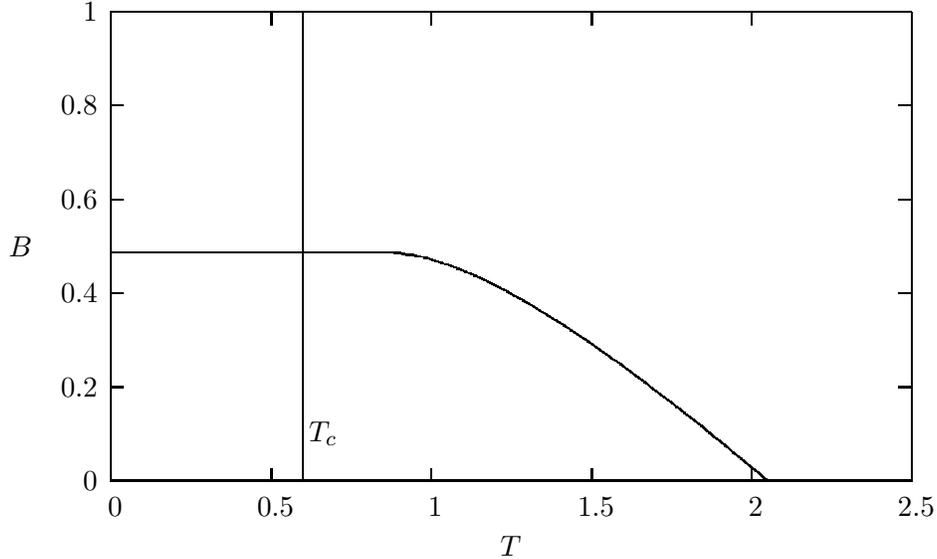

\include{barriera}
\caption{The free-energy barrier to relax from the ground state
$\vec\s^*$.}
\label{fig:barriera}
\end{figure}

We can understand better the structure of the excitations studying the
window free-energy as a function of the overlap $q$, i.e.\ the
free-energy of configurations that do not coincide with $\vec\s^*$ on
exaclty $d$ sites among the $\ell$ of the window, with $q=1-d/\ell$.
For simplicity we can consider the contribution of ``one bubble
configurations'' where all the $d$ sites in question are contiguous.
We show below that configurations with more that one bubble are
exponentially unprobable for large $m$ values.

For large $m$ the dominant contribution to the free-energy per link
$f(\beta,\ell,d)=\lim_{m\to\infty}F/m(\ell+1)$ is independent of the
position of the bubble and reads, for $d=1,\ldots,\ell$:
\begin{equation}
f(\beta,\ell,d) =\left\{
\begin{array}{cl}
-\frac{\ell-d}{\ell+1}\sqrt{\log(2)}-\frac{d+1}{\ell+1}
\left(\frac{\beta}{4}+\frac{T}{1+1/d}\log(2)\right)&
\quad T > T_c \sqrt{{1+1/d}}\;,\\
-\frac{\ell-d}{\ell+1}\sqrt{\log(2)}-\frac{d+1}{\ell+1}
\sqrt{\frac{1}{1+1/d}\log(2)}& \quad T \le T_c \sqrt{{1+1/d}}\;.
\end{array}
\right.
\label{eq:free2} 
\end{equation}
For $d=0$ the free-energy is simply given by
$f(\beta,\ell,0)=\epsilon^*$.  As one can explicitly see, $f$ is
monotonically decreasing in $d$: the completely open configuration is
always the most favoured among the ones with $d\geq 1$.  Notice that
at low temperature $f(\beta,\ell,1)>f(\beta,\ell,0)$ and the
difference $B(\beta)=(l+1)\big[f(\beta,\ell,1)-f(\beta,\ell,0)\big]$
can be interpreted as a relaxation free-energy barrier for a system
prepared in the ground state $\vec\s^*$. The barrier $B(\beta)$,
plotted in Fig.~\ref{fig:barriera}, is $\ell$-independent and vanishes
at a temperature $T=T_c/(1-1/\sqrt{2})$.

If we remove the assumption of considering only one-bubble
configurations, the free-energy in Eq.(\ref{eq:free2}) becomes
\begin{equation}
f(\beta,\ell,d) = \min_{b\in\{1,d/2\}} f(\beta,\ell,d,b)\;,
\label{eq:freeBub}
\end{equation}
where $f(\beta,\ell,d,b)$ is the free-energy of configurations
differing in $d$ variables from $\vec\s^*$ and having $b$ bubbles,
given by the following expression in the large $m$ limit:
\begin{equation}
f(\beta,\ell,d,b) =\left\{
\begin{array}{cl}
-\frac{\ell+1-d-b}{\ell+1}\sqrt{\log(2)}-\frac{d+b}{\ell+1}
\left(\frac{\beta}{4}+\frac{T}{1+b/d}\log(2)\right)&
\quad T > T_c \sqrt{{1+b/d}}\;,\\
-\frac{\ell+1-d-b}{\ell+1}\sqrt{\log(2)}-\frac{d+b}{\ell+1}
\sqrt{\frac{1}{1+b/d}\log(2)}& \quad T \le T_c \sqrt{{1+b/d}}\;.
\end{array}
\right.
\end{equation}
It is easy to verify that the minimum in Eq.(\ref{eq:freeBub}) is
always achieved in $b=1$, i.e.\ on one-bubble configurations.
Multi-bubble configurations can only modify the corrections to the
leading behavior in $m$.

\subsection{Analysis of the Ground State}

We would like to present here some attempts to take into account
finite $m$ contributions. Corrections to the asymptotic result have
two sources: the correlations between the levels and sample-to-sample
fluctuations. Though we were not able to deal with the former, we
could analyze some of the latter.

Actually we derive some analytical results under 2 main
approximations, namely (i) energy levels are basically treated as
uncorrelated and (ii) the energy of the reference configuration (the
ground state energy $E^*$) is considered to be evenly distributed
among the links, each one having a local energy $m \epsilon^* = E^* /
L$ (please note that the entire system is made of $L$ links, while the
window had $\ell+1$ links).  We will see below that numerical evidence
shows that this is the case not too close to the boundaries $i = 0$
and $i = L$.

The distribution of $\epsilon^*$ is known for $L=1$, since in that
case $m \epsilon^*$ corresponds to the minimum among $2^m$ independent
random Gaussian variables of variance $m/2$, that is
\begin{equation}
\epsilon^*(L=1) \eqd -\sqrt{\log(2)} +
\frac{\log(m) + \log(4\pi\log(2)) + 2 X}{4 m \sqrt{\log(2)}} +
\O\left(\frac{\log(m)}{m^2}\right)\;,
\end{equation}
where $X$ is a Gumbel distributed variable, i.e.\ $\P(X > x) =
e^{-e^x}$. Similarly a closed formula can be obtained for $L=2$ which
corresponds to a two level GREM.  Unfortunately as soon as $L>2$ there
are no exact results on the ground state energy of the model.  In this
case an analitical upper bound can be simply constructed by the
following greedy algorithm: given that $\s_0$ is fixed, assign $\s_1$
to the value minimizing $E(\s_0,\s_1)$ and repeat the procedure
recursively on the next variable; at each step the link energy has the
same probability distribution as $m \epsilon^*(L=1)$, and so the
global ground state energy satisfies $E^*(m,L) \le m L
\epsilon^*(L=1)$.

\begin{figure}
\includegraphics[width=.8\textwidth]{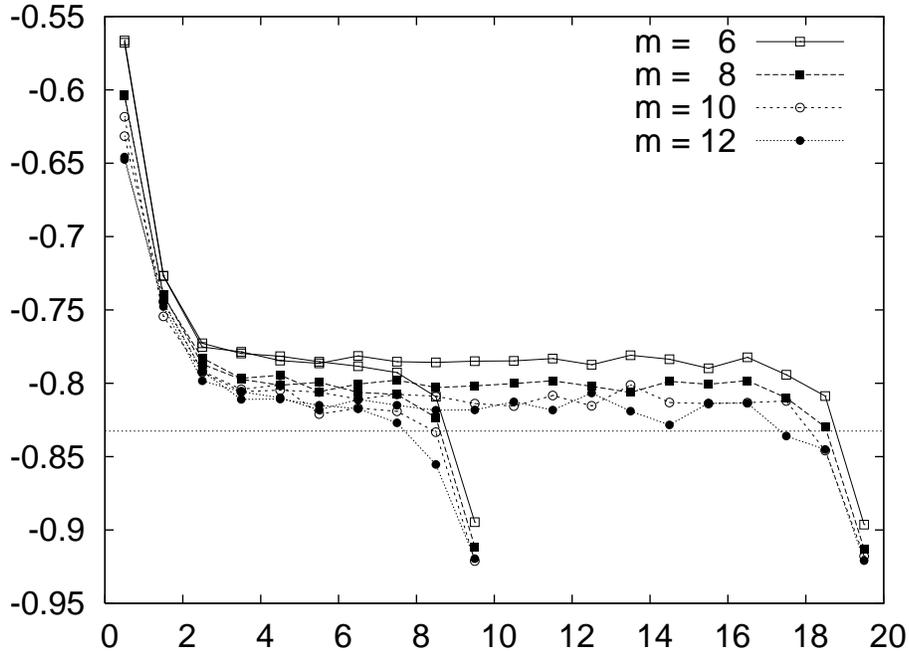}
\caption{The mean value of the ground state link energy as a function
  of the position in the system for $L=10$ and $L=20$.  On the left
  boundary the configuration is fixed, while on the right end the
  system is free.  The horizontal line is $-\sqrt{\log(2)}$.}
\label{fig:GSlocal}
\end{figure}

Our numerical data suggest this bound to be tight at the leading order
in $m$ for any value of $L$.  More precisely, we find numerically that
the mean ground state energy can be very well fitted, for large values
of $L$, by the following formula:
\begin{equation}
\frac{E^*(m,L)}{m\,L} \simeq -\sqrt{\log(2)} + \frac{A}{m^{3/2}} +
\frac{B}{m^{1/2} L}\;,
\label{eq:EGS}
\end{equation}
with $A \sim 0.7$ and $B \sim 0.4$.  This behavior clearly shows that
the convergence to the asymptotic intensive energy, $-\sqrt{\log(2)}$,
becomes faster increasing $L$: for $L=1$ corrections are
$\O(\frac{\log(m)}{m})$ and they become $\O(m^{-3/2})$ in the
$L\to\infty$ limit. We see from Fig.~\ref{fig:GSlocal} that already
for $L\sim 10$, not too close to the boundaries, ground states link
energies are independent of $L$, and their numerical values are well
represented by the previous formula with $L=\infty$.  Obviuosly, given
the values of $m$ we can study, formula (\ref{eq:EGS}) has to be taken
as an empirical interpolating function.  We find from our data that in
ground state configurations, link energies have very small
sample-to-sample fluctuations, which decrease for larger $m$ values:
for this reason considering only mean values for the link energies is
a good approximation.  In order to minimize finite $L$ effects and
have a homogeneous ground state inside the window we find that it was
enough to consider sistem sizes $L=\ell+20$, i.e.\ 10 sites between
the window and system boundaries.

Once understood the ground state structure, let us now turn to the
estimate of the window correlation functions.

\subsection{Finite $m$ estimates of the correlation functions}

Under the assumptions stated above, the weight, in the window
partition function, of all the configurations differing in $d$
variables with respect to $\vec\s^*$ is given by
\begin{equation}
Z_d = (\ell+1-d)\,(2^m-1)^d\;\frac{
\int_{(d+1)m\epsilon^*}^\infty dz\; e^{-\beta z -\frac{z^2}{m(d+1)}}}
{\int_{(d+1)m\epsilon^*}^\infty dz\; e^{-\frac{z^2}{m(d+1)}}}\;
e^{-\beta(\ell-d)m\epsilon^*}\;,
\label{eq:Zd}
\end{equation}
where the first term gives the number of ways to place a bubble of
size $d$ in a window of size $\ell$, the second term counts the number
of configurations of the $d$ variables which have to differ from
$\vec\s^*$, the fraction is the average of $e^{-\beta H}$ over the
p.d.f.\ of the energies of the bubble (it is the sum of $d+1$ Gaussian
variables of variance $m/2$, bounded from below by the ground state
energy, $(d+1)m\epsilon^*$) and the last term is given by the $l-d$
links having the ground state energy.  In equation (\ref{eq:Zd}) we
have that $d \in \{1,\ldots,\ell\}$, while the weight of the ground
state is given by $Z_0 = \exp(-\beta E^*) =
\exp(-\beta(\ell+1)m\epsilon^*)$.  We do not write explicitly the
dependence of $Z_d$ on $\beta$, $\ell$, $m$ and $\epsilon^*$ in order
to keep the notation light.

$Z_d$ is an annealed approximation for the window partition function
at a fixed distance from the ground state.  Still, the fact that we
keep the dependence on $\epsilon^*$ explicit is important in order to
control some fluctuations: e.g.\ both $\log(p_0)$ and $\log(q_0)$ are
given by free-energy differences, where the dependence on $\epsilon^*$
is partially canceled out, and their average over $\epsilon^*$ can be
done without any approximation.  The two observables we are interested
in are indeed given by
\begin{equation}
p_0 = \frac{Z_0}{\sum_{d=0}^\ell Z_d}\;,\qquad
q_0 = \frac{\sum_{d=0}^\ell (1-d/\ell) Z_d}{\sum_{d=0}^\ell Z_d}\;,
\label{eq:q0p0}
\end{equation}
and can be easily computed by evaluating numerically the integrals in
the definition of $Z_d$, once the p.d.f.\ of $\epsilon^*$ is known.
We have measured numerically such a distribution, but once we plugged
it into Eq.(\ref{eq:Zd}) we discovered that the observables we are
interested in ($p_{typ}$ and $q_{typ}$) mainly depend on the mean of
$\epsilon^*$, being such a distribution very narrow.  Moreover we are
mostly interested in the dependence of these observables on $m$ in
order to understand the approach to the $m\to\infty$ limit, and the
average of $\epsilon^*$ carries the largest dependence on $m$.

For these reasons the analytical curves we are going to compare with
numerical data in the next section have been obtained using a
non-fluctuating value for $\epsilon^*$, give by Eq.(\ref{eq:EGS}) that
is $\epsilon^* = -\sqrt{\log(2)} + 0.726 / m ^{3/2}$.  As we show
below, this dependence on the interaction range is already enough to
produce strong finite $m$ effects.  Remind that, in the $m\to\infty$
limit, the logarithm of $Z_d$ is given by the free-energy in
Eq.(\ref{eq:free2}), and both $p_{typ}$ and $q_{typ}$ should drop from
1 to 0 when $\ell$ crosses the value of $\ell_c(\beta)$ given by
Eq.(\ref{eq:ell_c}).

\section{Numerical results}\label{sec4}

The aim of this section is to compute numerically the above defined
critical length scale for the 1D random energy model.  The numerical
experiment we have performed consists in:
\begin{compactenum}
\item computing the ground state of a system of size $L$;
\item fixing the ground state configuration outside a window of size
 $\ell$;
\item computing $q_{typ}$ and $p_{typ}$ in order to see whether there
  is a first order transition in these quantities varying the window
  size $\ell$.
\end{compactenum}
In the $m \to \infty$ limit we expect such a transition when the
window size crosses the value $\ell_c(\beta)$ given in
Eq.(\ref{eq:ell_c}).  For finite values of $m$ the system cannot have
any transition (it is one-dimensional), but still the crossover may be
very sharp.  Our main interest is in understanding how much the
behavior of finite $m$ systems resembles the mean-field (i.e. $m \to
\infty$) limit and how fast is the convergence.

As explained in the previous section, we take the size of the system
$L$ larger than the size of the window in order to avoid boundary
effects; that is, to all practical purposes we are working in the $L
\to \infty$ limit.

Thanks to the one-dimensional topology all the experiments can be done
exactly by tranfer matrix methods.  Unfortunately for each link we
have a different random matrix with $2^m \times 2^m$ entries; for this
reason we are forced to small values of $m$ (actually we use
$m=6,8,10,12$).  Please note that these $m$ values are not so small:
the number of degrees of freedom per region ($2^m$) is comparable or
even larger than the number of particles within a typical region
studied in realistic models of glassy systems \cite{Cavagna}.  Since
we are interested in computing the free-energy at a given value of the
overlap with the reference configuration, the transfer matrix
computation is slightly more complicated and requires a total time of
order $\mathcal{O}(\ell^2 2^{2m})$.  The average over the disorder is
done with at least 1000 samples for any $m$ value.

\begin{figure}
\includegraphics[width=\textwidth]{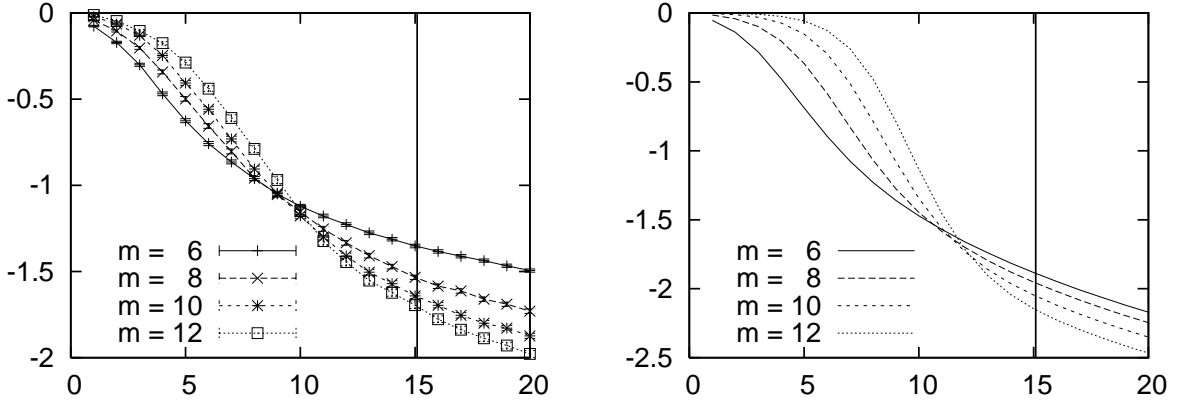}
\caption{The overlap $q_{typ}$ as a function of $\ell$ at
  $T=0.8$. Left panel: numerical values obtained throught the transfer
  matrix algorithm for various values of $m$. Right panel: analytic
  curves obtained through the approximations discussed in the
  text. The vertical lines marks the value of $\ell_c$ in the
  $m\to\infty$ limit. }
\label{fig:qtyp_T08}
\end{figure}

\begin{figure}
\includegraphics[width=\textwidth]{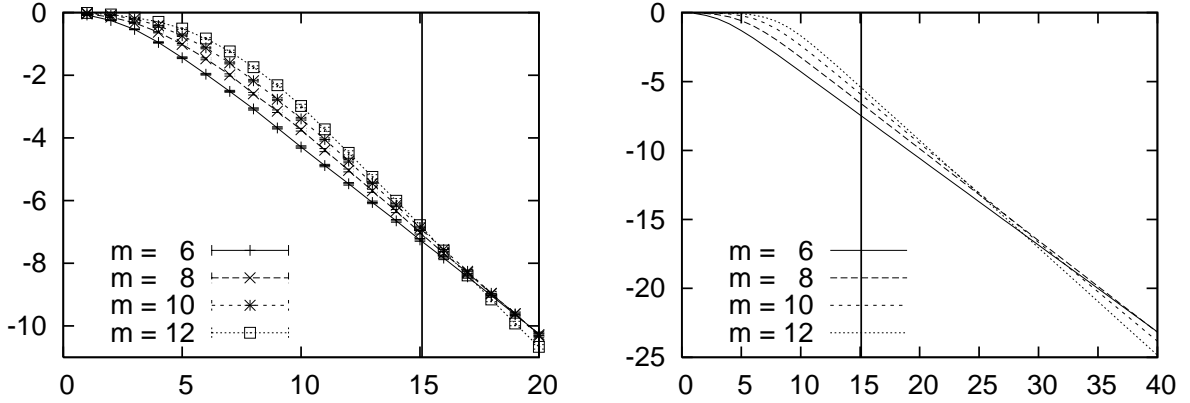}
\caption{Same as figure \ref{fig:qtyp_T08} for the quantity $p_{typ}$.}
\label{fig:ptyp_T08}
\end{figure}

We are going to present results for temperature $T=0.8$, which is a
very reasonable value (in the $m \to \infty$ limit the critical
temperature is $T_c=0.60056...$), since the critical window size
is $\ell_c(T=0.8) = 15.09$.

In Fig.~\ref{fig:qtyp_T08} and \ref{fig:ptyp_T08} we show respectiely
$\log(q_{typ})$ and $\log(p_{typ})$ as a a function of $\ell$.  Left
panels reports data from exact numerical computations, while right
panels show the outcome of the analytical approximated analysis.  The
vertical line is the critical window size $\ell_c$.

Some comments are in order.  The behavior of all the curves for
different $m$ hints at a first order transition for $m\to\infty$
separating a high overlap region at small $\ell$ from a zero overlap
region at large $\ell$. This behavior is in agreement with the
prediction of the mosaic theory, however, the convergence is very
slow! Indeed so slow, that it does not allow an estimate of the speed
of convergence.  The crossing point of numerical data for
$\log(q_{typ})$ is around $\ell = 10$, well below the predicted
$\ell_c=15.09$.  In principle one could argue that the one-dimensional
model may have a first order transition at a lower value of $\ell_c$,
but the crossing point of numerical data for $\log(p_{typ})$, taking
place around $\ell=17$, suggests that the crossing point is strongly
dependent on $m$ and converges for $m\to\infty$ somewhere between 10
and 17 (we are assuming that both $q_{typ}$ and $p_{typ}$ have a jump
at the same value of $\ell$ for $m\to\infty$).

Still more evident indications of strong finite $m$ effects come from
the analytical curves (see right panels of Fig.~\ref{fig:qtyp_T08} and
Fig.~\ref{fig:ptyp_T08}): these have been computed from
Eq.(\ref{eq:Zd}) and Eq.(\ref{eq:q0p0}) with $\epsilon^* =
-\sqrt{\log(2)} + 0.726\,m^{-3/2}$, see Eq.(\ref{eq:EGS}), which is the
best interpolation for the ground state energy in the window, far from
the boundaries.  Although these curves have been obtained under some
approximations, they look qualitatively very similar to the exact
numerical data, and also quantitatively are not far from the data.
For the analytical curves we know that they have a jump in $\ell =
\ell_c$ in the $m\to\infty$ limit, still for the present values of $m$
they show a crossing point quite far from $\ell_c$.

Moreover the value of the overlap at the crossing point may be very
small, depending on the overlap one is looking at (see e.g.\ the value
of $p_{typ}$ at the crossing point). For this reason may be very
difficult to locate the crossing point (remember that our model has a
very strong random first order transition in the $m\to\infty$ limit,
and most probably things work even worst in more realistic models!).

We remark that simulating the model for a single value of $m$ it would
be difficult to claim any agreement with the 1RSB theory of glasses
and the mosaic state: it is comparing different values on the
interaction range $m$ that the agreement becomes apparent. One could
argument that one dimension is the worst possibility to observe any
behavior reminiscent of a phase transition and in higher dimension the
situation could be more favourable to the theory. Recent simulations
of more realistic binary Lennard-Jones mixtures \cite{Cavagna}
however, failed to identify a sharp mosaic length.

\begin{figure}
\includegraphics[width=.8\textwidth]{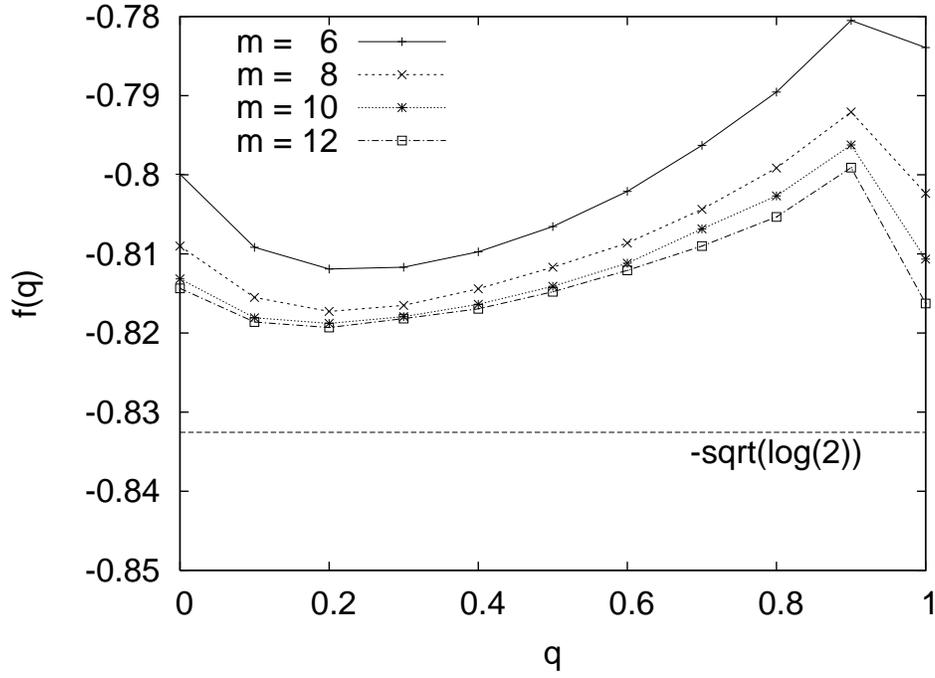}
\caption{Window free-energy for $\ell=10$ and $T=0.8$ as a function of
the overlap with respect to the ground state configuration.}
\label{fig:fq_T08}
\end{figure}

\begin{figure}
\includegraphics[width=.8\textwidth]{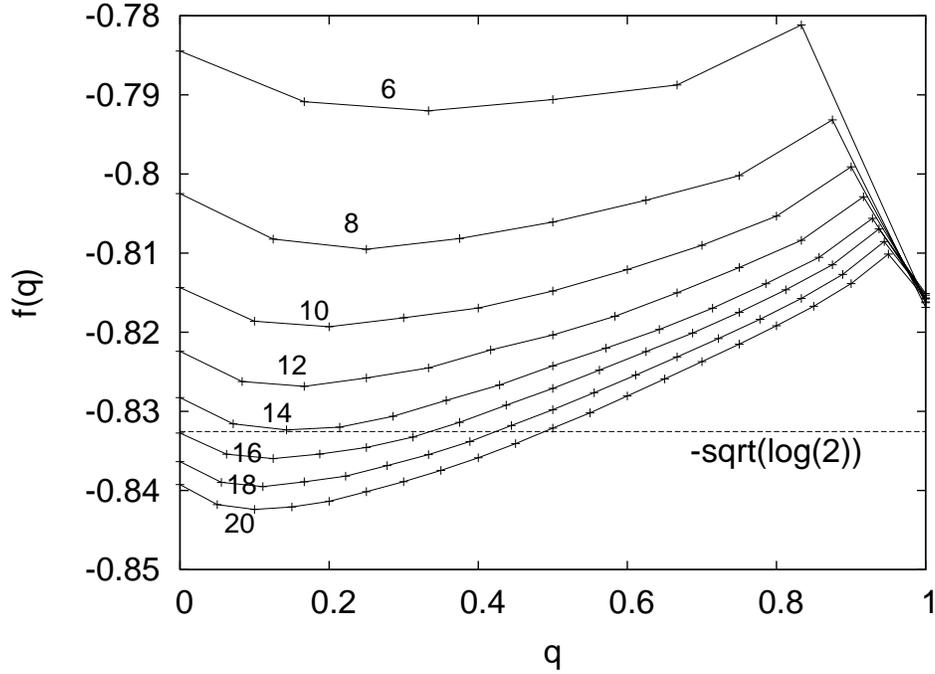}
\caption{Window free-energy for $T=0.8$, $m=12$ and many different
  values of $\ell$, from 6 (top) to 20 (bottom).}
\label{fig:Zq_T08}
\end{figure}

In order to understand better why $q_{typ}$ has such strong finite $m$
corrections and show an effective crossing point at window sizes
smaller than $\ell_c$, we have studied the window free energy as a
function of the overlap with respect to the ground state.  We show in
Fig.~\ref{fig:fq_T08} such a free-energy for $T=0.8$, $\ell=10$ and
many values of $m$ in order to study the dependence on $m$.  We see
that, increasing the value of $m$, all the curves $f(q)$ tends to
decrease, but corrections to the $m\to\infty$ limit are clearly larger
for $f(q=1)$ than for the rest of the curve.  Please note that
$f(q=1)$ corresponds to the ground state energy, that converges in the
$m\to\infty$ limit to $-\sqrt{\log(2)}$ (represented by the horizontal
line in the plot).  The different convergence rate for different $q$
values can be understood also from the analytical computation in the
previous Section; indeed in the expression for $Z_d$, see
Eq.(\ref{eq:Zd}), larger corrections are for small $d$ values
(corresponding to larger $q$).

A discrepancy with respect to the analytical computation, is that the
free-energy presents a minimum at a positive value of the overlap,
while in the $m\to\infty$ limit we expect the minimum to be in $q=0$.
This may be one more effect of the slow convergence to the mean-field
limit.

In the tentative of extrapolating the numerical results to the
$m\to\infty$ limit, we have fitted $f(q)$ data at fixed $q$, finding
that the limit of $f(q=1)$ is always compatible with
$-\sqrt{\log(2)}$, while for $q<1$ the asymptotic value of $f(q)$ is
quite close to that computed numerically with $m=12$, especially close
to the minimum of $f(q)$.

In Fig.~\ref{fig:Zq_T08} we show the free-energy $f(q)$ for $m=12$
(which is very close to the $m\to\infty$ value in the low $q$ region)
for many $\ell$ values, ranging from 6 to 20 (top to bottom).  The
apparent first order phase transition between the $q=1$ and the small
overlap regimes is taking place between $\ell=9$ and $\ell=10$ when
the minimum goes below $f(q=1)$~\footnote{Please note also that
$f(q=1)$ does not depend on the window size $\ell$, confirming that
the ground state in the window is insensitive to the boundaries.},
consistently to what we observe in the left panel of
Fig.~\ref{fig:qtyp_T08}.  Nonetheless, in the $m\to\infty$ we expect
the transition to take place when the minimum goes below the value
$-\sqrt{\log(2)}$, and we see from Fig.~\ref{fig:Zq_T08} that this
happens around $\ell=14$, much closer to the predicted
$\ell_c(T=0.8)=15.09$.

\section{Conclusions} 

The scope of this paper is to study the properties of convergence to
the mosaic picture in models with larger and larger interaction
ranges.  We showed that, as it should be expected, the behavior of
point to set correlations approach the behavior predicted by the
mosaic picture for large interaction range.  The numerical evidence in
favor of that comes from a differential analysis comparing the
behavior for different values of the interaction range $m$.  Curves at
single values of $m$ do not allow to distinguish mosaic behavior from
a single state picture where the point to set correlation exhibit a
smooth behavior as a function of $\ell$.  This is unfortunate as it
indicates that it could be difficult to find confirmations or
disprovals of the mosaic picture in realistic glass former models on
the basis of the behavior of point to set correlations.

Some papers have recently addressed the study of point-to-set
functions in non disordered models. Amazingly, the model where the
mosaic predictions seems to fit better the data is a kinetically
constrained model considered in \cite{garrahan} where a step like
behavior of the overlap as a function of the window size is observed.
Conversely, for a Lennard-Jones binary mixture, though it is observed
a characteristic length growing with temperature, no step behavior is
seen. We remark on this purpose that in our data it would been
difficult to decide in favor of the mosaic picture on the basis of a
single value of $m$. It is only comparing different values of $m$ that
evidence for the first order jump has been obtained.

Moreover the convergence to the large $m$ limit is rather slow: finite
$m$ curves are very smooth and show no precursor of the asymptotic
step-like behavior.  Ref.~\cite{Montanari} studied a similar 1d model
with finite interaction range, namely a XORSAT model. The main
difference with respect to our study is that the model studied in
ref.~\cite{Montanari} possesses zero-energy ground states and it has
been studied only at zero temperature.  Despite these differences also
in ref.~\cite{Montanari} large finite-range effects have been found.

The main effect that we have seen in the model studied here is that a
rather sharp transition takes place at a finite temperature
$\beta_c(\ell)$ between a single low-energy ground state (i.e.\ of
zero complexity) and a set of higher free-energy states (with positive
complexity) in a way more or less similar to the mosaic
picture. However this transition is plagued by large fluctuations
mainly due to the energy of the ground state, which plays a
fundamental role in determining the critical temperature: the final
effect being a sizable smoothing of the random first order phase
transition at finite value of the interaction range $m$.

The conclusions reached in this work suggest that the direct
observation of the phase transition predicted within the mosaic theory
may be rather difficult in realistic models, where the interaction
range cannot be made very large.  A smarter approach for the
identification of such a transition is likely needed.

\end{document}

%% file: parabole.tex
\setlength{\unitlength}{0.240900pt}
\ifx\plotpoint\undefined\newsavebox{\plotpoint}\fi
\begin{picture}(1500,900)(0,0)
\sbox{\plotpoint}{\rule[-0.200pt]{0.400pt}{0.400pt}}%
\put(181.0,123.0){\rule[-0.200pt]{4.818pt}{0.400pt}}
\put(161,123){\makebox(0,0)[r]{ 0}}
\put(1419.0,123.0){\rule[-0.200pt]{4.818pt}{0.400pt}}
\put(181.0,228.0){\rule[-0.200pt]{4.818pt}{0.400pt}}
\put(161,228){\makebox(0,0)[r]{ 0.1}}
\put(1419.0,228.0){\rule[-0.200pt]{4.818pt}{0.400pt}}
\put(181.0,334.0){\rule[-0.200pt]{4.818pt}{0.400pt}}
\put(161,334){\makebox(0,0)[r]{ 0.2}}
\put(1419.0,334.0){\rule[-0.200pt]{4.818pt}{0.400pt}}
\put(181.0,439.0){\rule[-0.200pt]{4.818pt}{0.400pt}}
\put(161,439){\makebox(0,0)[r]{ 0.3}}
\put(1419.0,439.0){\rule[-0.200pt]{4.818pt}{0.400pt}}
\put(181.0,544.0){\rule[-0.200pt]{4.818pt}{0.400pt}}
\put(161,544){\makebox(0,0)[r]{ 0.4}}
\put(1419.0,544.0){\rule[-0.200pt]{4.818pt}{0.400pt}}
\put(181.0,649.0){\rule[-0.200pt]{4.818pt}{0.400pt}}
\put(161,649){\makebox(0,0)[r]{ 0.5}}
\put(1419.0,649.0){\rule[-0.200pt]{4.818pt}{0.400pt}}
\put(181.0,755.0){\rule[-0.200pt]{4.818pt}{0.400pt}}
\put(161,755){\makebox(0,0)[r]{ 0.6}}
\put(1419.0,755.0){\rule[-0.200pt]{4.818pt}{0.400pt}}
\put(181.0,860.0){\rule[-0.200pt]{4.818pt}{0.400pt}}
\put(161,860){\makebox(0,0)[r]{ 0.7}}
\put(1419.0,860.0){\rule[-0.200pt]{4.818pt}{0.400pt}}
\put(181.0,123.0){\rule[-0.200pt]{0.400pt}{4.818pt}}
\put(181,82){\makebox(0,0){-1}}
\put(181.0,840.0){\rule[-0.200pt]{0.400pt}{4.818pt}}
\put(496.0,123.0){\rule[-0.200pt]{0.400pt}{4.818pt}}
\put(496,82){\makebox(0,0){-0.5}}
\put(496.0,840.0){\rule[-0.200pt]{0.400pt}{4.818pt}}
\put(810.0,123.0){\rule[-0.200pt]{0.400pt}{4.818pt}}
\put(810,82){\makebox(0,0){ 0}}
\put(810.0,840.0){\rule[-0.200pt]{0.400pt}{4.818pt}}
\put(1125.0,123.0){\rule[-0.200pt]{0.400pt}{4.818pt}}
\put(1125,82){\makebox(0,0){ 0.5}}
\put(1125.0,840.0){\rule[-0.200pt]{0.400pt}{4.818pt}}
\put(1439.0,123.0){\rule[-0.200pt]{0.400pt}{4.818pt}}
\put(1439,82){\makebox(0,0){ 1}}
\put(1439.0,840.0){\rule[-0.200pt]{0.400pt}{4.818pt}}
\put(181.0,123.0){\rule[-0.200pt]{303.052pt}{0.400pt}}
\put(1439.0,123.0){\rule[-0.200pt]{0.400pt}{177.543pt}}
\put(181.0,860.0){\rule[-0.200pt]{303.052pt}{0.400pt}}
\put(181.0,123.0){\rule[-0.200pt]{0.400pt}{177.543pt}}
\put(40,491){\makebox(0,0){$S_\ell(\epsilon)$}}
\put(810,21){\makebox(0,0){$\epsilon$}}
\put(1125,649){\makebox(0,0)[l]{$\ell = \infty$}}
\put(1018,544){\makebox(0,0)[r]{$\ell = 4$}}
\put(286,81){\makebox(0,0){$\epsilon^*$}}
\put(389,207){\makebox(0,0){\large $\searrow$}}
\put(420,165){\makebox(0,0)[l]{slope $\beta_c(\ell=4)$}}
\multiput(286.58,123.00)(0.500,0.769){753}{\rule{0.120pt}{0.715pt}}
\multiput(285.17,123.00)(378.000,579.516){2}{\rule{0.400pt}{0.357pt}}
\sbox{\plotpoint}{\rule[-0.500pt]{1.000pt}{1.000pt}}%
\multiput(286,123)(7.030,19.529){2}{\usebox{\plotpoint}}
\multiput(295,148)(7.413,19.387){2}{\usebox{\plotpoint}}
\put(315.16,200.73){\usebox{\plotpoint}}
\multiput(321,216)(7.093,19.506){2}{\usebox{\plotpoint}}
\multiput(333,249)(8.027,19.141){2}{\usebox{\plotpoint}}
\put(353.23,297.24){\usebox{\plotpoint}}
\multiput(359,311)(8.253,19.044){2}{\usebox{\plotpoint}}
\put(377.44,354.61){\usebox{\plotpoint}}
\multiput(384,371)(8.740,18.825){2}{\usebox{\plotpoint}}
\put(402.96,411.38){\usebox{\plotpoint}}
\multiput(410,426)(8.430,18.967){2}{\usebox{\plotpoint}}
\put(429.38,467.75){\usebox{\plotpoint}}
\put(438.77,486.26){\usebox{\plotpoint}}
\multiput(448,504)(9.885,18.250){2}{\usebox{\plotpoint}}
\put(467.93,541.27){\usebox{\plotpoint}}
\put(477.98,559.42){\usebox{\plotpoint}}
\put(488.54,577.29){\usebox{\plotpoint}}
\multiput(499,595)(10.679,17.798){2}{\usebox{\plotpoint}}
\put(521.01,630.41){\usebox{\plotpoint}}
\put(532.63,647.61){\usebox{\plotpoint}}
\put(544.62,664.55){\usebox{\plotpoint}}
\put(556.67,681.45){\usebox{\plotpoint}}
\put(568.99,698.14){\usebox{\plotpoint}}
\put(582.12,714.21){\usebox{\plotpoint}}
\put(595.35,730.19){\usebox{\plotpoint}}
\put(609.06,745.76){\usebox{\plotpoint}}
\put(624.00,760.15){\usebox{\plotpoint}}
\put(638.78,774.72){\usebox{\plotpoint}}
\put(654.27,788.51){\usebox{\plotpoint}}
\put(670.97,800.83){\usebox{\plotpoint}}
\put(687.74,813.05){\usebox{\plotpoint}}
\put(705.45,823.86){\usebox{\plotpoint}}
\put(723.86,833.43){\usebox{\plotpoint}}
\put(743.17,840.97){\usebox{\plotpoint}}
\put(763.01,847.08){\usebox{\plotpoint}}
\put(783.39,850.83){\usebox{\plotpoint}}
\put(804.02,853.00){\usebox{\plotpoint}}
\put(824.75,852.33){\usebox{\plotpoint}}
\put(845.30,849.45){\usebox{\plotpoint}}
\put(865.40,844.49){\usebox{\plotpoint}}
\put(885.12,838.03){\usebox{\plotpoint}}
\put(904.01,829.49){\usebox{\plotpoint}}
\put(922.16,819.44){\usebox{\plotpoint}}
\put(939.30,807.77){\usebox{\plotpoint}}
\put(956.25,795.80){\usebox{\plotpoint}}
\put(972.44,782.83){\usebox{\plotpoint}}
\put(987.47,768.53){\usebox{\plotpoint}}
\put(1002.47,754.18){\usebox{\plotpoint}}
\put(1016.93,739.31){\usebox{\plotpoint}}
\put(1030.14,723.32){\usebox{\plotpoint}}
\put(1043.65,707.56){\usebox{\plotpoint}}
\put(1056.35,691.15){\usebox{\plotpoint}}
\put(1068.41,674.26){\usebox{\plotpoint}}
\put(1080.53,657.41){\usebox{\plotpoint}}
\put(1092.34,640.35){\usebox{\plotpoint}}
\put(1103.78,623.03){\usebox{\plotpoint}}
\put(1114.75,605.41){\usebox{\plotpoint}}
\put(1125.38,587.59){\usebox{\plotpoint}}
\multiput(1134,573)(10.559,-17.869){2}{\usebox{\plotpoint}}
\put(1156.14,533.47){\usebox{\plotpoint}}
\put(1165.95,515.18){\usebox{\plotpoint}}
\put(1175.71,496.86){\usebox{\plotpoint}}
\multiput(1185,479)(9.282,-18.564){2}{\usebox{\plotpoint}}
\put(1203.31,441.06){\usebox{\plotpoint}}
\multiput(1210,426)(9.004,-18.701){2}{\usebox{\plotpoint}}
\put(1229.66,384.65){\usebox{\plotpoint}}
\multiput(1236,371)(7.708,-19.271){2}{\usebox{\plotpoint}}
\put(1253.92,327.33){\usebox{\plotpoint}}
\multiput(1261,311)(8.027,-19.141){2}{\usebox{\plotpoint}}
\multiput(1274,280)(8.027,-19.141){2}{\usebox{\plotpoint}}
\put(1293.41,231.38){\usebox{\plotpoint}}
\multiput(1299,216)(7.413,-19.387){2}{\usebox{\plotpoint}}
\multiput(1312,182)(7.413,-19.387){2}{\usebox{\plotpoint}}
\put(1329.95,134.25){\usebox{\plotpoint}}
\put(1334,123){\usebox{\plotpoint}}
\sbox{\plotpoint}{\rule[-0.600pt]{1.200pt}{1.200pt}}%
\put(341.51,123){\rule{1.200pt}{2.650pt}}
\multiput(339.51,123.00)(4.000,5.500){2}{\rule{1.200pt}{1.325pt}}
\multiput(348.24,134.00)(0.501,1.192){16}{\rule{0.121pt}{3.162pt}}
\multiput(343.51,134.00)(13.000,24.438){2}{\rule{1.200pt}{1.581pt}}
\multiput(361.24,165.00)(0.501,1.151){16}{\rule{0.121pt}{3.069pt}}
\multiput(356.51,165.00)(13.000,23.630){2}{\rule{1.200pt}{1.535pt}}
\multiput(374.24,195.00)(0.501,1.254){14}{\rule{0.121pt}{3.300pt}}
\multiput(369.51,195.00)(12.000,23.151){2}{\rule{1.200pt}{1.650pt}}
\multiput(386.24,225.00)(0.501,1.068){16}{\rule{0.121pt}{2.885pt}}
\multiput(381.51,225.00)(13.000,22.013){2}{\rule{1.200pt}{1.442pt}}
\multiput(399.24,253.00)(0.501,1.027){16}{\rule{0.121pt}{2.792pt}}
\multiput(394.51,253.00)(13.000,21.204){2}{\rule{1.200pt}{1.396pt}}
\multiput(412.24,280.00)(0.501,1.119){14}{\rule{0.121pt}{3.000pt}}
\multiput(407.51,280.00)(12.000,20.773){2}{\rule{1.200pt}{1.500pt}}
\multiput(424.24,307.00)(0.501,0.986){16}{\rule{0.121pt}{2.700pt}}
\multiput(419.51,307.00)(13.000,20.396){2}{\rule{1.200pt}{1.350pt}}
\multiput(437.24,333.00)(0.501,0.945){16}{\rule{0.121pt}{2.608pt}}
\multiput(432.51,333.00)(13.000,19.588){2}{\rule{1.200pt}{1.304pt}}
\multiput(450.24,358.00)(0.501,0.904){16}{\rule{0.121pt}{2.515pt}}
\multiput(445.51,358.00)(13.000,18.779){2}{\rule{1.200pt}{1.258pt}}
\multiput(463.24,382.00)(0.501,0.939){14}{\rule{0.121pt}{2.600pt}}
\multiput(458.51,382.00)(12.000,17.604){2}{\rule{1.200pt}{1.300pt}}
\multiput(475.24,405.00)(0.501,0.822){16}{\rule{0.121pt}{2.331pt}}
\multiput(470.51,405.00)(13.000,17.162){2}{\rule{1.200pt}{1.165pt}}
\multiput(488.24,427.00)(0.501,0.822){16}{\rule{0.121pt}{2.331pt}}
\multiput(483.51,427.00)(13.000,17.162){2}{\rule{1.200pt}{1.165pt}}
\multiput(501.24,449.00)(0.501,0.849){14}{\rule{0.121pt}{2.400pt}}
\multiput(496.51,449.00)(12.000,16.019){2}{\rule{1.200pt}{1.200pt}}
\multiput(513.24,470.00)(0.501,0.698){16}{\rule{0.121pt}{2.054pt}}
\multiput(508.51,470.00)(13.000,14.737){2}{\rule{1.200pt}{1.027pt}}
\multiput(526.24,489.00)(0.501,0.698){16}{\rule{0.121pt}{2.054pt}}
\multiput(521.51,489.00)(13.000,14.737){2}{\rule{1.200pt}{1.027pt}}
\multiput(539.24,508.00)(0.501,0.657){16}{\rule{0.121pt}{1.962pt}}
\multiput(534.51,508.00)(13.000,13.929){2}{\rule{1.200pt}{0.981pt}}
\multiput(552.24,526.00)(0.501,0.669){14}{\rule{0.121pt}{2.000pt}}
\multiput(547.51,526.00)(12.000,12.849){2}{\rule{1.200pt}{1.000pt}}
\multiput(564.24,543.00)(0.501,0.616){16}{\rule{0.121pt}{1.869pt}}
\multiput(559.51,543.00)(13.000,13.120){2}{\rule{1.200pt}{0.935pt}}
\multiput(577.24,560.00)(0.501,0.534){16}{\rule{0.121pt}{1.685pt}}
\multiput(572.51,560.00)(13.000,11.503){2}{\rule{1.200pt}{0.842pt}}
\multiput(590.24,575.00)(0.501,0.579){14}{\rule{0.121pt}{1.800pt}}
\multiput(585.51,575.00)(12.000,11.264){2}{\rule{1.200pt}{0.900pt}}
\multiput(602.24,590.00)(0.501,0.493){16}{\rule{0.121pt}{1.592pt}}
\multiput(597.51,590.00)(13.000,10.695){2}{\rule{1.200pt}{0.796pt}}
\multiput(613.00,606.24)(0.489,0.501){14}{\rule{1.600pt}{0.121pt}}
\multiput(613.00,601.51)(9.679,12.000){2}{\rule{0.800pt}{1.200pt}}
\multiput(628.24,616.00)(0.501,0.489){14}{\rule{0.121pt}{1.600pt}}
\multiput(623.51,616.00)(12.000,9.679){2}{\rule{1.200pt}{0.800pt}}
\multiput(638.00,631.24)(0.533,0.502){12}{\rule{1.718pt}{0.121pt}}
\multiput(638.00,626.51)(9.434,11.000){2}{\rule{0.859pt}{1.200pt}}
\multiput(651.00,642.24)(0.587,0.502){10}{\rule{1.860pt}{0.121pt}}
\multiput(651.00,637.51)(9.139,10.000){2}{\rule{0.930pt}{1.200pt}}
\multiput(664.00,652.24)(0.651,0.502){8}{\rule{2.033pt}{0.121pt}}
\multiput(664.00,647.51)(8.780,9.000){2}{\rule{1.017pt}{1.200pt}}
\multiput(677.00,661.24)(0.588,0.502){8}{\rule{1.900pt}{0.121pt}}
\multiput(677.00,656.51)(8.056,9.000){2}{\rule{0.950pt}{1.200pt}}
\multiput(689.00,670.24)(0.732,0.503){6}{\rule{2.250pt}{0.121pt}}
\multiput(689.00,665.51)(8.330,8.000){2}{\rule{1.125pt}{1.200pt}}
\multiput(702.00,678.24)(0.835,0.505){4}{\rule{2.529pt}{0.122pt}}
\multiput(702.00,673.51)(7.752,7.000){2}{\rule{1.264pt}{1.200pt}}
\multiput(715.00,685.24)(0.792,0.509){2}{\rule{2.700pt}{0.123pt}}
\multiput(715.00,680.51)(6.396,6.000){2}{\rule{1.350pt}{1.200pt}}
\put(727,689.01){\rule{3.132pt}{1.200pt}}
\multiput(727.00,686.51)(6.500,5.000){2}{\rule{1.566pt}{1.200pt}}
\put(740,693.51){\rule{3.132pt}{1.200pt}}
\multiput(740.00,691.51)(6.500,4.000){2}{\rule{1.566pt}{1.200pt}}
\put(753,697.51){\rule{3.132pt}{1.200pt}}
\multiput(753.00,695.51)(6.500,4.000){2}{\rule{1.566pt}{1.200pt}}
\put(766,700.51){\rule{2.891pt}{1.200pt}}
\multiput(766.00,699.51)(6.000,2.000){2}{\rule{1.445pt}{1.200pt}}
\put(778,702.51){\rule{3.132pt}{1.200pt}}
\multiput(778.00,701.51)(6.500,2.000){2}{\rule{1.566pt}{1.200pt}}
\put(791,704.01){\rule{3.132pt}{1.200pt}}
\multiput(791.00,703.51)(6.500,1.000){2}{\rule{1.566pt}{1.200pt}}
\put(816,704.01){\rule{3.132pt}{1.200pt}}
\multiput(816.00,704.51)(6.500,-1.000){2}{\rule{1.566pt}{1.200pt}}
\put(829,702.51){\rule{3.132pt}{1.200pt}}
\multiput(829.00,703.51)(6.500,-2.000){2}{\rule{1.566pt}{1.200pt}}
\put(842,700.51){\rule{2.891pt}{1.200pt}}
\multiput(842.00,701.51)(6.000,-2.000){2}{\rule{1.445pt}{1.200pt}}
\put(854,697.51){\rule{3.132pt}{1.200pt}}
\multiput(854.00,699.51)(6.500,-4.000){2}{\rule{1.566pt}{1.200pt}}
\put(867,693.51){\rule{3.132pt}{1.200pt}}
\multiput(867.00,695.51)(6.500,-4.000){2}{\rule{1.566pt}{1.200pt}}
\put(880,689.01){\rule{3.132pt}{1.200pt}}
\multiput(880.00,691.51)(6.500,-5.000){2}{\rule{1.566pt}{1.200pt}}
\multiput(893.00,686.25)(0.792,-0.509){2}{\rule{2.700pt}{0.123pt}}
\multiput(893.00,686.51)(6.396,-6.000){2}{\rule{1.350pt}{1.200pt}}
\multiput(905.00,680.26)(0.835,-0.505){4}{\rule{2.529pt}{0.122pt}}
\multiput(905.00,680.51)(7.752,-7.000){2}{\rule{1.264pt}{1.200pt}}
\multiput(918.00,673.26)(0.732,-0.503){6}{\rule{2.250pt}{0.121pt}}
\multiput(918.00,673.51)(8.330,-8.000){2}{\rule{1.125pt}{1.200pt}}
\multiput(931.00,665.26)(0.588,-0.502){8}{\rule{1.900pt}{0.121pt}}
\multiput(931.00,665.51)(8.056,-9.000){2}{\rule{0.950pt}{1.200pt}}
\multiput(943.00,656.26)(0.651,-0.502){8}{\rule{2.033pt}{0.121pt}}
\multiput(943.00,656.51)(8.780,-9.000){2}{\rule{1.017pt}{1.200pt}}
\multiput(956.00,647.26)(0.587,-0.502){10}{\rule{1.860pt}{0.121pt}}
\multiput(956.00,647.51)(9.139,-10.000){2}{\rule{0.930pt}{1.200pt}}
\multiput(969.00,637.26)(0.533,-0.502){12}{\rule{1.718pt}{0.121pt}}
\multiput(969.00,637.51)(9.434,-11.000){2}{\rule{0.859pt}{1.200pt}}
\multiput(984.24,622.36)(0.501,-0.489){14}{\rule{0.121pt}{1.600pt}}
\multiput(979.51,625.68)(12.000,-9.679){2}{\rule{1.200pt}{0.800pt}}
\multiput(994.00,613.26)(0.489,-0.501){14}{\rule{1.600pt}{0.121pt}}
\multiput(994.00,613.51)(9.679,-12.000){2}{\rule{0.800pt}{1.200pt}}
\multiput(1009.24,597.39)(0.501,-0.493){16}{\rule{0.121pt}{1.592pt}}
\multiput(1004.51,600.70)(13.000,-10.695){2}{\rule{1.200pt}{0.796pt}}
\multiput(1022.24,582.53)(0.501,-0.579){14}{\rule{0.121pt}{1.800pt}}
\multiput(1017.51,586.26)(12.000,-11.264){2}{\rule{1.200pt}{0.900pt}}
\multiput(1034.24,568.01)(0.501,-0.534){16}{\rule{0.121pt}{1.685pt}}
\multiput(1029.51,571.50)(13.000,-11.503){2}{\rule{1.200pt}{0.842pt}}
\multiput(1047.24,552.24)(0.501,-0.616){16}{\rule{0.121pt}{1.869pt}}
\multiput(1042.51,556.12)(13.000,-13.120){2}{\rule{1.200pt}{0.935pt}}
\multiput(1060.24,534.70)(0.501,-0.669){14}{\rule{0.121pt}{2.000pt}}
\multiput(1055.51,538.85)(12.000,-12.849){2}{\rule{1.200pt}{1.000pt}}
\multiput(1072.24,517.86)(0.501,-0.657){16}{\rule{0.121pt}{1.962pt}}
\multiput(1067.51,521.93)(13.000,-13.929){2}{\rule{1.200pt}{0.981pt}}
\multiput(1085.24,499.47)(0.501,-0.698){16}{\rule{0.121pt}{2.054pt}}
\multiput(1080.51,503.74)(13.000,-14.737){2}{\rule{1.200pt}{1.027pt}}
\multiput(1098.24,480.47)(0.501,-0.698){16}{\rule{0.121pt}{2.054pt}}
\multiput(1093.51,484.74)(13.000,-14.737){2}{\rule{1.200pt}{1.027pt}}
\multiput(1111.24,460.04)(0.501,-0.849){14}{\rule{0.121pt}{2.400pt}}
\multiput(1106.51,465.02)(12.000,-16.019){2}{\rule{1.200pt}{1.200pt}}
\multiput(1123.24,439.32)(0.501,-0.822){16}{\rule{0.121pt}{2.331pt}}
\multiput(1118.51,444.16)(13.000,-17.162){2}{\rule{1.200pt}{1.165pt}}
\multiput(1136.24,417.32)(0.501,-0.822){16}{\rule{0.121pt}{2.331pt}}
\multiput(1131.51,422.16)(13.000,-17.162){2}{\rule{1.200pt}{1.165pt}}
\multiput(1149.24,394.21)(0.501,-0.939){14}{\rule{0.121pt}{2.600pt}}
\multiput(1144.51,399.60)(12.000,-17.604){2}{\rule{1.200pt}{1.300pt}}
\multiput(1161.24,371.56)(0.501,-0.904){16}{\rule{0.121pt}{2.515pt}}
\multiput(1156.51,376.78)(13.000,-18.779){2}{\rule{1.200pt}{1.258pt}}
\multiput(1174.24,347.18)(0.501,-0.945){16}{\rule{0.121pt}{2.608pt}}
\multiput(1169.51,352.59)(13.000,-19.588){2}{\rule{1.200pt}{1.304pt}}
\multiput(1187.24,321.79)(0.501,-0.986){16}{\rule{0.121pt}{2.700pt}}
\multiput(1182.51,327.40)(13.000,-20.396){2}{\rule{1.200pt}{1.350pt}}
\multiput(1200.24,294.55)(0.501,-1.119){14}{\rule{0.121pt}{3.000pt}}
\multiput(1195.51,300.77)(12.000,-20.773){2}{\rule{1.200pt}{1.500pt}}
\multiput(1212.24,268.41)(0.501,-1.027){16}{\rule{0.121pt}{2.792pt}}
\multiput(1207.51,274.20)(13.000,-21.204){2}{\rule{1.200pt}{1.396pt}}
\multiput(1225.24,241.03)(0.501,-1.068){16}{\rule{0.121pt}{2.885pt}}
\multiput(1220.51,247.01)(13.000,-22.013){2}{\rule{1.200pt}{1.442pt}}
\multiput(1238.24,211.30)(0.501,-1.254){14}{\rule{0.121pt}{3.300pt}}
\multiput(1233.51,218.15)(12.000,-23.151){2}{\rule{1.200pt}{1.650pt}}
\multiput(1250.24,182.26)(0.501,-1.151){16}{\rule{0.121pt}{3.069pt}}
\multiput(1245.51,188.63)(13.000,-23.630){2}{\rule{1.200pt}{1.535pt}}
\multiput(1263.24,151.88)(0.501,-1.192){16}{\rule{0.121pt}{3.162pt}}
\multiput(1258.51,158.44)(13.000,-24.438){2}{\rule{1.200pt}{1.581pt}}
\put(1273.51,123){\rule{1.200pt}{2.650pt}}
\multiput(1271.51,128.50)(4.000,-5.500){2}{\rule{1.200pt}{1.325pt}}
\put(804.0,707.0){\rule[-0.600pt]{2.891pt}{1.200pt}}
\sbox{\plotpoint}{\rule[-0.400pt]{0.800pt}{0.800pt}}%
\put(286,123){\circle*{18}}
\sbox{\plotpoint}{\rule[-0.200pt]{0.400pt}{0.400pt}}%
\put(181.0,123.0){\rule[-0.200pt]{303.052pt}{0.400pt}}
\put(1439.0,123.0){\rule[-0.200pt]{0.400pt}{177.543pt}}
\put(181.0,860.0){\rule[-0.200pt]{303.052pt}{0.400pt}}
\put(181.0,123.0){\rule[-0.200pt]{0.400pt}{177.543pt}}
\end{picture}

%% file: barriera.tex
\setlength{\unitlength}{0.240900pt}
\ifx\plotpoint\undefined\newsavebox{\plotpoint}\fi
\begin{picture}(1500,900)(0,0)
\sbox{\plotpoint}{\rule[-0.200pt]{0.400pt}{0.400pt}}%
\put(181.0,123.0){\rule[-0.200pt]{4.818pt}{0.400pt}}
\put(161,123){\makebox(0,0)[r]{ 0}}
\put(1419.0,123.0){\rule[-0.200pt]{4.818pt}{0.400pt}}
\put(181.0,270.0){\rule[-0.200pt]{4.818pt}{0.400pt}}
\put(161,270){\makebox(0,0)[r]{ 0.2}}
\put(1419.0,270.0){\rule[-0.200pt]{4.818pt}{0.400pt}}
\put(181.0,418.0){\rule[-0.200pt]{4.818pt}{0.400pt}}
\put(161,418){\makebox(0,0)[r]{ 0.4}}
\put(1419.0,418.0){\rule[-0.200pt]{4.818pt}{0.400pt}}
\put(181.0,565.0){\rule[-0.200pt]{4.818pt}{0.400pt}}
\put(161,565){\makebox(0,0)[r]{ 0.6}}
\put(1419.0,565.0){\rule[-0.200pt]{4.818pt}{0.400pt}}
\put(181.0,713.0){\rule[-0.200pt]{4.818pt}{0.400pt}}
\put(161,713){\makebox(0,0)[r]{ 0.8}}
\put(1419.0,713.0){\rule[-0.200pt]{4.818pt}{0.400pt}}
\put(181.0,860.0){\rule[-0.200pt]{4.818pt}{0.400pt}}
\put(161,860){\makebox(0,0)[r]{ 1}}
\put(1419.0,860.0){\rule[-0.200pt]{4.818pt}{0.400pt}}
\put(181.0,123.0){\rule[-0.200pt]{0.400pt}{4.818pt}}
\put(181,82){\makebox(0,0){ 0}}
\put(181.0,840.0){\rule[-0.200pt]{0.400pt}{4.818pt}}
\put(433.0,123.0){\rule[-0.200pt]{0.400pt}{4.818pt}}
\put(433,82){\makebox(0,0){ 0.5}}
\put(433.0,840.0){\rule[-0.200pt]{0.400pt}{4.818pt}}
\put(684.0,123.0){\rule[-0.200pt]{0.400pt}{4.818pt}}
\put(684,82){\makebox(0,0){ 1}}
\put(684.0,840.0){\rule[-0.200pt]{0.400pt}{4.818pt}}
\put(936.0,123.0){\rule[-0.200pt]{0.400pt}{4.818pt}}
\put(936,82){\makebox(0,0){ 1.5}}
\put(936.0,840.0){\rule[-0.200pt]{0.400pt}{4.818pt}}
\put(1187.0,123.0){\rule[-0.200pt]{0.400pt}{4.818pt}}
\put(1187,82){\makebox(0,0){ 2}}
\put(1187.0,840.0){\rule[-0.200pt]{0.400pt}{4.818pt}}
\put(1439.0,123.0){\rule[-0.200pt]{0.400pt}{4.818pt}}
\put(1439,82){\makebox(0,0){ 2.5}}
\put(1439.0,840.0){\rule[-0.200pt]{0.400pt}{4.818pt}}
\put(181.0,123.0){\rule[-0.200pt]{303.052pt}{0.400pt}}
\put(1439.0,123.0){\rule[-0.200pt]{0.400pt}{177.543pt}}
\put(181.0,860.0){\rule[-0.200pt]{303.052pt}{0.400pt}}
\put(181.0,123.0){\rule[-0.200pt]{0.400pt}{177.543pt}}
\put(40,491){\makebox(0,0){$B$}}
\put(810,21){\makebox(0,0){$T$}}
\put(493,197){\makebox(0,0)[l]{$T_c$}}
\put(483,123){\line(0,1){737}}
\put(181,482){\usebox{\plotpoint}}
\put(626,480.17){\rule{2.500pt}{0.400pt}}
\multiput(626.00,481.17)(6.811,-2.000){2}{\rule{1.250pt}{0.400pt}}
\put(638,478.17){\rule{2.700pt}{0.400pt}}
\multiput(638.00,479.17)(7.396,-2.000){2}{\rule{1.350pt}{0.400pt}}
\put(651,476.17){\rule{2.700pt}{0.400pt}}
\multiput(651.00,477.17)(7.396,-2.000){2}{\rule{1.350pt}{0.400pt}}
\multiput(664.00,474.95)(2.695,-0.447){3}{\rule{1.833pt}{0.108pt}}
\multiput(664.00,475.17)(9.195,-3.000){2}{\rule{0.917pt}{0.400pt}}
\multiput(677.00,471.94)(1.651,-0.468){5}{\rule{1.300pt}{0.113pt}}
\multiput(677.00,472.17)(9.302,-4.000){2}{\rule{0.650pt}{0.400pt}}
\multiput(689.00,467.94)(1.797,-0.468){5}{\rule{1.400pt}{0.113pt}}
\multiput(689.00,468.17)(10.094,-4.000){2}{\rule{0.700pt}{0.400pt}}
\multiput(702.00,463.94)(1.797,-0.468){5}{\rule{1.400pt}{0.113pt}}
\multiput(702.00,464.17)(10.094,-4.000){2}{\rule{0.700pt}{0.400pt}}
\multiput(715.00,459.93)(1.267,-0.477){7}{\rule{1.060pt}{0.115pt}}
\multiput(715.00,460.17)(9.800,-5.000){2}{\rule{0.530pt}{0.400pt}}
\multiput(727.00,454.93)(1.378,-0.477){7}{\rule{1.140pt}{0.115pt}}
\multiput(727.00,455.17)(10.634,-5.000){2}{\rule{0.570pt}{0.400pt}}
\multiput(740.00,449.93)(1.123,-0.482){9}{\rule{0.967pt}{0.116pt}}
\multiput(740.00,450.17)(10.994,-6.000){2}{\rule{0.483pt}{0.400pt}}
\multiput(753.00,443.93)(1.378,-0.477){7}{\rule{1.140pt}{0.115pt}}
\multiput(753.00,444.17)(10.634,-5.000){2}{\rule{0.570pt}{0.400pt}}
\multiput(766.00,438.93)(0.874,-0.485){11}{\rule{0.786pt}{0.117pt}}
\multiput(766.00,439.17)(10.369,-7.000){2}{\rule{0.393pt}{0.400pt}}
\multiput(778.00,431.93)(1.123,-0.482){9}{\rule{0.967pt}{0.116pt}}
\multiput(778.00,432.17)(10.994,-6.000){2}{\rule{0.483pt}{0.400pt}}
\multiput(791.00,425.93)(0.950,-0.485){11}{\rule{0.843pt}{0.117pt}}
\multiput(791.00,426.17)(11.251,-7.000){2}{\rule{0.421pt}{0.400pt}}
\multiput(804.00,418.93)(0.874,-0.485){11}{\rule{0.786pt}{0.117pt}}
\multiput(804.00,419.17)(10.369,-7.000){2}{\rule{0.393pt}{0.400pt}}
\multiput(816.00,411.93)(0.950,-0.485){11}{\rule{0.843pt}{0.117pt}}
\multiput(816.00,412.17)(11.251,-7.000){2}{\rule{0.421pt}{0.400pt}}
\multiput(829.00,404.93)(0.950,-0.485){11}{\rule{0.843pt}{0.117pt}}
\multiput(829.00,405.17)(11.251,-7.000){2}{\rule{0.421pt}{0.400pt}}
\multiput(842.00,397.93)(0.758,-0.488){13}{\rule{0.700pt}{0.117pt}}
\multiput(842.00,398.17)(10.547,-8.000){2}{\rule{0.350pt}{0.400pt}}
\multiput(854.00,389.93)(0.824,-0.488){13}{\rule{0.750pt}{0.117pt}}
\multiput(854.00,390.17)(11.443,-8.000){2}{\rule{0.375pt}{0.400pt}}
\multiput(867.00,381.93)(0.824,-0.488){13}{\rule{0.750pt}{0.117pt}}
\multiput(867.00,382.17)(11.443,-8.000){2}{\rule{0.375pt}{0.400pt}}
\multiput(880.00,373.93)(0.824,-0.488){13}{\rule{0.750pt}{0.117pt}}
\multiput(880.00,374.17)(11.443,-8.000){2}{\rule{0.375pt}{0.400pt}}
\multiput(893.00,365.93)(0.758,-0.488){13}{\rule{0.700pt}{0.117pt}}
\multiput(893.00,366.17)(10.547,-8.000){2}{\rule{0.350pt}{0.400pt}}
\multiput(905.00,357.93)(0.728,-0.489){15}{\rule{0.678pt}{0.118pt}}
\multiput(905.00,358.17)(11.593,-9.000){2}{\rule{0.339pt}{0.400pt}}
\multiput(918.00,348.93)(0.824,-0.488){13}{\rule{0.750pt}{0.117pt}}
\multiput(918.00,349.17)(11.443,-8.000){2}{\rule{0.375pt}{0.400pt}}
\multiput(931.00,340.93)(0.669,-0.489){15}{\rule{0.633pt}{0.118pt}}
\multiput(931.00,341.17)(10.685,-9.000){2}{\rule{0.317pt}{0.400pt}}
\multiput(943.00,331.93)(0.728,-0.489){15}{\rule{0.678pt}{0.118pt}}
\multiput(943.00,332.17)(11.593,-9.000){2}{\rule{0.339pt}{0.400pt}}
\multiput(956.00,322.93)(0.728,-0.489){15}{\rule{0.678pt}{0.118pt}}
\multiput(956.00,323.17)(11.593,-9.000){2}{\rule{0.339pt}{0.400pt}}
\multiput(969.00,313.93)(0.728,-0.489){15}{\rule{0.678pt}{0.118pt}}
\multiput(969.00,314.17)(11.593,-9.000){2}{\rule{0.339pt}{0.400pt}}
\multiput(982.00,304.93)(0.669,-0.489){15}{\rule{0.633pt}{0.118pt}}
\multiput(982.00,305.17)(10.685,-9.000){2}{\rule{0.317pt}{0.400pt}}
\multiput(994.00,295.92)(0.652,-0.491){17}{\rule{0.620pt}{0.118pt}}
\multiput(994.00,296.17)(11.713,-10.000){2}{\rule{0.310pt}{0.400pt}}
\multiput(1007.00,285.93)(0.728,-0.489){15}{\rule{0.678pt}{0.118pt}}
\multiput(1007.00,286.17)(11.593,-9.000){2}{\rule{0.339pt}{0.400pt}}
\multiput(1020.00,276.92)(0.600,-0.491){17}{\rule{0.580pt}{0.118pt}}
\multiput(1020.00,277.17)(10.796,-10.000){2}{\rule{0.290pt}{0.400pt}}
\multiput(1032.00,266.92)(0.652,-0.491){17}{\rule{0.620pt}{0.118pt}}
\multiput(1032.00,267.17)(11.713,-10.000){2}{\rule{0.310pt}{0.400pt}}
\multiput(1045.00,256.93)(0.728,-0.489){15}{\rule{0.678pt}{0.118pt}}
\multiput(1045.00,257.17)(11.593,-9.000){2}{\rule{0.339pt}{0.400pt}}
\multiput(1058.00,247.92)(0.600,-0.491){17}{\rule{0.580pt}{0.118pt}}
\multiput(1058.00,248.17)(10.796,-10.000){2}{\rule{0.290pt}{0.400pt}}
\multiput(1070.00,237.92)(0.652,-0.491){17}{\rule{0.620pt}{0.118pt}}
\multiput(1070.00,238.17)(11.713,-10.000){2}{\rule{0.310pt}{0.400pt}}
\multiput(1083.00,227.92)(0.652,-0.491){17}{\rule{0.620pt}{0.118pt}}
\multiput(1083.00,228.17)(11.713,-10.000){2}{\rule{0.310pt}{0.400pt}}
\multiput(1096.00,217.92)(0.652,-0.491){17}{\rule{0.620pt}{0.118pt}}
\multiput(1096.00,218.17)(11.713,-10.000){2}{\rule{0.310pt}{0.400pt}}
\multiput(1109.00,207.92)(0.543,-0.492){19}{\rule{0.536pt}{0.118pt}}
\multiput(1109.00,208.17)(10.887,-11.000){2}{\rule{0.268pt}{0.400pt}}
\multiput(1121.00,196.92)(0.652,-0.491){17}{\rule{0.620pt}{0.118pt}}
\multiput(1121.00,197.17)(11.713,-10.000){2}{\rule{0.310pt}{0.400pt}}
\multiput(1134.00,186.92)(0.652,-0.491){17}{\rule{0.620pt}{0.118pt}}
\multiput(1134.00,187.17)(11.713,-10.000){2}{\rule{0.310pt}{0.400pt}}
\multiput(1147.00,176.92)(0.543,-0.492){19}{\rule{0.536pt}{0.118pt}}
\multiput(1147.00,177.17)(10.887,-11.000){2}{\rule{0.268pt}{0.400pt}}
\multiput(1159.00,165.92)(0.652,-0.491){17}{\rule{0.620pt}{0.118pt}}
\multiput(1159.00,166.17)(11.713,-10.000){2}{\rule{0.310pt}{0.400pt}}
\multiput(1172.00,155.92)(0.590,-0.492){19}{\rule{0.573pt}{0.118pt}}
\multiput(1172.00,156.17)(11.811,-11.000){2}{\rule{0.286pt}{0.400pt}}
\multiput(1185.00,144.92)(0.652,-0.491){17}{\rule{0.620pt}{0.118pt}}
\multiput(1185.00,145.17)(11.713,-10.000){2}{\rule{0.310pt}{0.400pt}}
\multiput(1198.00,134.92)(0.543,-0.492){19}{\rule{0.536pt}{0.118pt}}
\multiput(1198.00,135.17)(10.887,-11.000){2}{\rule{0.268pt}{0.400pt}}
\put(1210,123.17){\rule{0.700pt}{0.400pt}}
\multiput(1210.00,124.17)(1.547,-2.000){2}{\rule{0.350pt}{0.400pt}}
\put(181.0,482.0){\rule[-0.200pt]{107.200pt}{0.400pt}}
\put(181.0,123.0){\rule[-0.200pt]{303.052pt}{0.400pt}}
\put(1439.0,123.0){\rule[-0.200pt]{0.400pt}{177.543pt}}
\put(181.0,860.0){\rule[-0.200pt]{303.052pt}{0.400pt}}
\put(181.0,123.0){\rule[-0.200pt]{0.400pt}{177.543pt}}
\end{picture}